# A Reduced-Boundary-Function Method for Convective Heat Transfer with Axial Heat Conduction and Viscous Dissipation

Zhijie Xu[1,a]

1. Energy Resource Recovery & Management, Idaho National Laboratory, Idaho Falls, Idaho 83415 USA. Now at

Computational Mathematics Group, Fundamental and Computational Sciences Directorate, Pacific Northwest National Laboratory, Richland, WA 99352, USA

## Abstract

We introduce a new method of solution for the convective heat transfer under forced laminar flow that is confined by two parallel plates with a distance of 2*a* or by a circular tube with a radius of *a*. The advection-conduction equation is first mapped onto the boundary. The original problem of solving the unknown field $T(x,r,t)$ is reduced to seek the solutions of *T* at the boundary (*r=a* or *r=0*, *r* is the distance from the centerline shown in Fig. 1), i.e. the boundary functions $T_a(x,t) \equiv T(x,r=a,t)$ and/or $T_0(x,t) \equiv T(x,r=0,t)$. In this manner, the original problem is significantly simplified by reducing the problem dimensionality from 3 to 2. The unknown field $T(x,r,t)$ can be eventually solved in terms of these boundary functions. The method is applied to the convective heat transfer with uniform wall temperature boundary condition and with heat exchange between flowing fluids and its surroundings that is relevant to the geothermal applications. Analytical solutions are presented and validated for the steady state problem using the proposed method.

[a] Electronic mail: zhijie.xu@pnnl.gov, zhijiexu@hotmail.com, Tel: 509-372-4885

## 1. Introduction

Convective heat transfer in circular tube and flat plates is of fundamental interest and practical importance and attracts considerable attention in the past. The analysis of convective heat transfer in tube was originally studied by Graetz [1, 2] with assumptions of steady and incompressible flow with fully developed velocity profile, negligible viscous dissipation, axial conduction, and temperature independent fluid properties. These assumptions were made to simplify the complexity while solving the heat transport equation. A steady-state flow of fluid with a constant temperature $T_i$ passing through a channel is shown in Fig.1. The wall boundary is prescribed with a uniform temperature $T_w$ in the original work of Graetz [1, 2] and Nusselt [3], and the model has been referred to as the Graetz problem that represents a class of simplified convective heat transfer problems.

A number of studies have been carried out on various extensions of the original Graetz problem for different applications. For example, the so-called "extended Graetz problem" refers to the including of axial conduction in the original Graetz problem and has been studied extensively [4-8]. The Peclet number (the dimensionless ratio of convective to conductive heat transfer) in the flow could be sufficiently large in many practical applications so that the axial heat conduction can be neglected. However, it is not the case for the heat transfer with small Peclet numbers, for example, in compact heat exchangers where liquid metals are used as the working fluids and axial conduction can play an important role [9].

The thermal-fluid system becomes increasingly smaller with the rapid growth of semiconductor technology. The technological applications in such applications as micro-electromechanical systems (MEMS) require more efficient convective heat transfer in a relatively smaller space at the micro-scale level. As the size of flow channel is reduced, the convective heat transfer does not follow the standard results based on the continuum assumption, which leads to the extension of the original Graetz problem to include the slip flow boundary and to allow the viscous dissipation and temperature jump at the wall [10-12].

The specific objective of this paper is to present a method of solution to the convective heat transfer with applications to two different boundary conditions. The example problems consider the effects of both axial conduction and viscous dissipation. Previous attempts to solve Graetz problem and its extensions are mostly taking the way of plugging a solution of series expansion into the advection-conduction equation. The method of solution presented in this paper uses a reduced-boundary-function approach, which significantly simplifies the original problem and reduces the problem dimensionality without losing numerical accuracy.

Moving boundary problems are important and numerically difficulty to solve since boundary or interface location is not known. Lagrangian particle methods have been developed to solve such problems without explicit interface tracking/capturing [13-15]. The method of solution presented in this paper can be applied to these problems in order to reduce the problem dimensionality. Examples are the solute precipitation and dissolution due to the chemical reaction [16, 17] and crystal growth. It was also applied to the thermal oxidation problem [18], where an oxidized layer is constantly growing due to the oxidation reaction.

In the problem of convective heat transfer, the advection-conduction equation is mapped onto the boundary, and the original problem to find temperature solution $T(x,r,t)$ is reduced to find the temperature solutions at the boundary $T_a(x,t) \equiv T(x, r=a, t)$ and/or $T_0(x,t) \equiv T(x, r=0, t)$. The original problem is significantly simplified with this method because the dimensionality of the problem is reduced from 3 to 2. In this paper, we introduce the formulation of the reduced boundary function method in Section 2, followed by the analytical, numerical solutions, and comparison for two boundary conditions in Sections 3 and 4.

## 2. The Formulation of the Reduced-Boundary-Function Method

The rigorous mathematical formulation is based on the advection-conduction equation for the convective heat transfer. The problem considered here is that a fluid is flowing with a steady Poiseuille flow through a channel confined by two flat plates or a circular tube. The fluid enters the channel at a prescribed temperature $T_i$. Two types of boundary conditions are considered. First one is a uniform wall temperature $T_w$ applied on the flow boundary. The second one is a mixed Dirichlet and Neumann boundary to mimic the heat exchange between the flowing fluid and the surrounding environment. The physical parameters are assumed to be independent of temperature. The geometry of the problem is defined in Fig. 1. The dynamics of temperature field $T$ is governed by the advection-conduction equation in the cylindrical coordinates,

$$\frac{\partial T}{\partial t}+v(r)\frac{\partial T}{\partial x}=D\left(\frac{\partial^2 T}{\partial r^2}+\frac{d}{r}\frac{\partial T}{\partial r}+\frac{\partial^2 T}{\partial x^2}\right)+\frac{\mu}{\rho c_p}\left(\frac{\partial v}{\partial r}\right)^2, \tag{1}$$

where $T(x,r,t)$ is the unknown temperature field depending on the axial position $x$, radial position $r$ and time $t$. $D=k/\rho c_p$ is the thermal diffusivity that is independent of the temperature $T$, where $k$ is the fluid thermal conductivity. $\rho$ is the density of fluid and $c_p$ is the heat capacity. The flow velocity profile $v(r)$ is varying in the radial direction, but is independent of $x$ along the flow direction. A steady-state flow is assumed to be fully developed and established. $d$ is a dimensionless number to indicate the flow is confined by two flat plates ($d = 0$) or by a circular tube ($d = 1$). The last term on the RHS (Right Hand Side) of Eq. (1) is the viscous dissipation, where $\mu$ is the dynamic viscosity. The relevant boundary conditions for $T$ are,

$$\left.\frac{\partial T}{\partial r}\right|_{r=0} = 0, \tag{2}$$

Obviously, the velocity $v(r)$ also satisfy

$$v|_{r=a} = 0 \text{ and } \left.\frac{\partial v}{\partial r}\right|_{r=0} = 0. \tag{3}$$

The condition $\left(\partial v/\partial r\right)|_{r=0} = 0$ is from the symmetry of the flow. The parabolic velocity profile for Poiseuille flow is,

$$v(r) = u_0\left(1 - \frac{r^2}{a^2}\right), \tag{4}$$

where $u_0$ is the maximum velocity at $r=0$ and the mean velocity $\bar{u}$ is written as

$$\bar{u} = \frac{2u_0}{3+d}. \tag{5}$$

An additional relation can be derived from the l'Hôpital's rule in calculus and will be used later,

$$\left.\left(\frac{1}{r}\frac{\partial T}{\partial r}\right)\right|_{r=0} = \left.\frac{\partial^2 T}{\partial r^2}\right|_{r=0}. \tag{6}$$

Let $T_a(x,t) \equiv T(x, r=a, t)$ represents the temperature along the wall ($r=a$) and $T_0(x,t) \equiv T(x, r=0, t)$ is the temperature along the centerline ($r=0$). The essential idea of reduced boundary function method is to find solutions for boundary functions $T_0(x,t)$ and/or $T_a(x,t)$, instead of finding solutions for $T(x,r,t)$ in the original problem. The original problem can be significantly simplified with this method because the dimensionality of the problem is reduced from 3 to 2.

Since Eq. (1) must be satisfied everywhere including the boundary $r=a$ and $r=0$, the following relationship can be obtained from Eqs. (1)-(6) for boundary functions $T_a$ and $T_0$,

$$\frac{\partial T_a}{\partial t} = D\left[\frac{\partial^2 T}{\partial r^2}\bigg|_{r=a} + \frac{d}{a}\frac{\partial T}{\partial r}\bigg|_{r=a} + \frac{\partial^2 T_a}{\partial x^2}\right] + \frac{4\mu u_0^2}{\rho c_p a^2}, \tag{7}$$

$$\frac{\partial T_0}{\partial t} + u_0 \frac{\partial T_0}{\partial x} = D\left[\frac{\partial^2 T_0}{\partial x^2} + (d+1)\frac{\partial^2 T}{\partial r^2}\bigg|_{r=0}\right]. \tag{8}$$

The same set of equations can be rewritten as

$$\frac{\partial T_a}{\partial t} = D\left[T_{2,a} + \frac{d}{a}T_{1,a} + \frac{\partial^2 T_a}{\partial x^2}\right] + A, \tag{9}$$

$$\frac{\partial T_0}{\partial t} + u_0 \frac{\partial T_0}{\partial x} = D\left[\frac{\partial^2 T_0}{\partial x^2} + (d+1)T_{2,0}\right], \tag{10}$$

by introducing short notations to represent derivatives at $r=0$ and $r=a$,

$$T_{n,0} = \frac{\partial^n T}{\partial r^n}\bigg|_{r=0}, \quad T_{n,a} = \frac{\partial^n T}{\partial r^n}\bigg|_{r=a}, \quad \text{and } A = \frac{4\mu u_0^2}{\rho c_p a^2}. \tag{11}$$

In principle, the temperature $T$ can be expressed in terms of these derivatives through a Taylor expansion,

$$T(r,x,t) = T_0(x,t) + \sum_{n=2}^{\infty} \frac{r^n}{n!} T_{n,0}\big|_{r=0}, \tag{12}$$

where $n$ starts from 2 because $(\partial T/\partial r)\big|_{r=0} = 0$. The following relation can be easily verified with expansion (12),

$$\left( \frac{d}{r} \frac{\partial^2 T}{\partial r^2} - \frac{d}{r^2} \frac{\partial T}{\partial r} \right)\bigg|_{r=0} = \frac{d}{2} \frac{\partial^3 T}{\partial r^3}\bigg|_{r=0}. \tag{13}$$

By taking derivative of the governing equation (1) with respect to $r$, we have

$$\frac{\partial^2 T}{\partial t \partial r} + v(r) \frac{\partial^2 T}{\partial x \partial r} + \frac{\partial v}{\partial r} \frac{\partial T}{\partial x} = D \left( \frac{\partial^3 T}{\partial r^3} + \frac{d}{r} \frac{\partial^2 T}{\partial r^2} - \frac{d}{r^2} \frac{\partial T}{\partial r} + \frac{\partial^3 T}{\partial x^2 \partial r} \right) + \frac{2Ar}{a^2}. \tag{14}$$

Again, Eq. (14) can be mapped onto the boundary $r=a$ and $r=0$ and the following relationships can be established for boundary functions $T_{1,a}$ and $T_{3,0}$ with the help of Eq.(13),

$$\frac{\partial T_{1,a}}{\partial t} + \frac{\partial v}{\partial r}\bigg|_{r=a} \frac{\partial T_a}{\partial x} = D \left[ \frac{\partial^2 T_{1,a}}{\partial x^2} + T_{3,a} + \frac{d}{a} T_{2,a} - \frac{d}{a^2} T_{1,a} \right] + \frac{2A}{a}, \tag{15}$$

$$0 = D T_{3,0}. \tag{16}$$

More generally, it was found that

$$T_{2n+1,0} = 0 \text{ for } n = 0,1,2,3... \tag{17}$$

due to the symmetry of $T$ with respect to $r$. Up to now, we do not use any temperature boundary conditions at wall boundary. Therefore, Eqs. (9)-(10) and Eqs. (15)-(17) are valid for any arbitrary temperature boundary conditions. Additionally, we do not make any assumptions, such as negligible viscous dissipation, small axial conduction, which are common in order for an analytical solution.

## 3. First Type of Boundary -- The Uniform Wall Temperature $T_w$

First, we consider the standard Graetz problem, where the wall of the channel or the tube is maintained at a uniform temperature $T_w$ so that

$$T_a = T_w, \tag{18}$$

and

$$\frac{\partial T_a}{\partial t} = \frac{\partial T_a}{\partial x} = \frac{\partial^2 T_a}{\partial x^2} = 0. \tag{19}$$

Eq. (9) relating the boundary functions $T_{1,a}$ and $T_{2,a}$ is simplified to

$$0 = D\left[T_{2,a} + \frac{d}{a}T_{1,a}\right] + A. \tag{20}$$

We first use expansion (12) with terms up to the fourth order to approximate the temperature field $T$, namely

$$T(x,r,t) \approx T_0(x,t) + \frac{r^2}{2}T_{2,0}(x,t) + \frac{r^4}{24}T_{4,0}(x,t). \tag{21}$$

The following relationships from the expansion can be easily found,

$$T_{1,a} = aT_{2,0} + \frac{a^3}{6}T_{4,0}, \tag{22}$$

$$T_{2,a} = T_{2,0} + \frac{a^2}{2}T_{4,0}. \tag{23}$$

This can be rewritten in the matrix form,

$$\begin{bmatrix} T_{1,a} \\ T_{2,a} \end{bmatrix} = B\begin{bmatrix} T_{2,0} \\ T_{4,0} \end{bmatrix} = \begin{bmatrix} a & a^3/6 \\ 1 & a^2/2 \end{bmatrix}\begin{bmatrix} T_{2,0} \\ T_{4,0} \end{bmatrix}. \tag{24}$$

By inversing the matrix $B$ and using Eq. (20), we can solve for $T_{2,0}$ and $T_{4,0}$,

$$T_{2,0} = \frac{3+d}{2a}T_{1,a} + \frac{A}{2D} \text{ and } T_{4,0} = -\frac{3(1+d)}{a^3}T_{1,a} - \frac{3A}{Da^2}. \tag{25}$$

The uniform wall temperature boundary should be satisfied, where

$$T_a = T_w = T_0 + \frac{a^2}{2}T_{2,0} + \frac{a^4}{24}T_{4,0}. \tag{26}$$

Substitution of Eq. (25) into Eq. (26) leads to the relationship

$$T_w - T_0 = \frac{5+d}{8} T_{1,a} a + \frac{Aa^2}{8D}, \tag{27}$$

and

$$T_{2,0} = \frac{3+d}{5+d} \cdot \frac{4}{a^2} (T_w - T_0) + \frac{1}{5+d} \cdot \frac{A}{D}. \tag{28}$$

Substitution of Eq. (28) into Eq. (10) yields the final governing equation for the centerline temperature $T_0$,

$$\frac{\partial T_0}{\partial t} + u_0 \frac{\partial T_0}{\partial x} = D \frac{\partial^2 T_0}{\partial x^2} + \frac{(d+1)(d+3)}{(d+5)} \cdot \frac{4D}{a^2} (T_w - T_0) + A \cdot \frac{1+d}{5+d}. \tag{29}$$

It is shown from Eq. (29) that $T_0$ follows an advection-diffusion-reaction equation. The steady-state solution can be easily found for $T_0$,

$$T_0 = \left( T_i - T_w - \frac{Aa^2}{4D(3+d)} \right) e^{-\beta_1 \xi} + T_w + \frac{Aa^2}{4D(3+d)}, \tag{30}$$

where it is assumed that $T_0(x=0) = T_i$ at the inlet. $\xi = x/a$ is the normalized axial coordinate. The dimensionless number $\beta_1$ defines the characteristic length of temperature decay and is a function of dimensionless Peclet number $p_e$,

$$\beta_1 = \frac{p_e}{2} \left[ \sqrt{1 + \frac{16(d+1)(d+3)}{(d+5) p_e^2}} - 1 \right], \tag{31}$$

where $p_e = u_0 a/D$ represents the ratio between the convective and conductive heat transfer. With the assumption of negligible axial conduction ($p_e \to \infty$), the dimensionless number $\beta_1$ can be expressed as

$$\beta_1 \approx \frac{4}{p_e} \frac{(d+1)(d+3)}{(d+5)} \text{ for } p_e \to \infty, \tag{32}$$

which is infinite at small $p_e$. This is not physical and means that neglecting axial conduction can lead to significant errors at small $p_e$. The solution of boundary function $T_{1,a}$ can be obtained from Eq. (27),

$$T_{1,a} = -\frac{8}{(5+d)}\left(T_i - T_w - \frac{Aa^2}{4D(3+d)}\right)e^{-\beta_1\xi} - \frac{Aa^2}{D(3+d)}. \tag{33}$$

Instead of using expansion with terms up to the fourth order, we can try the solution with terms up to the sixth order to see the difference of solutions,

$$T(x,r,t) \approx T_0(x,t) + \frac{r^2}{2}T_{2,0} + \frac{r^4}{24}T_{4,0} + \frac{r^6}{720}T_{6,0}. \tag{34}$$

Equation (15) was not used before and must be used here because we have an additional unknown boundary function $T_{6,0}$ in the new expansion (34). Equation (15) can be simplified to (due to the uniform wall temperature $T_a = T_w$),

$$\frac{\partial T_{1,a}}{\partial t} = D\left[\frac{\partial^2 T_{1,a}}{\partial x^2} + T_{3,a} + \frac{d}{a}T_{2,a} - \frac{d}{a^2}T_{1,a}\right] + \frac{2A}{a}, \tag{35}$$

The matrix form of the relationship between $T_{n,a}$ and $T_{n,0}$ ($n$ = 1, 2, 3) is

$$\begin{bmatrix} T_{1,a} \\ T_{2,a} \\ T_{3,a} \end{bmatrix} = \begin{bmatrix} a & a^3/6 & a^5/120 \\ 1 & a^2/2 & a^4/24 \\ 0 & a & a^3/6 \end{bmatrix} \begin{bmatrix} T_{2,0} \\ T_{4,0} \\ T_{6,0} \end{bmatrix}. \tag{36}$$

With the help of Eqs. (20), (35), (36) and the boundary condition $T(x,a,t) = T_w$, the equations for boundary functions are

$$T_w - T_0 = \frac{a^3}{48D}\frac{\partial T_{1,a}}{\partial t} - \frac{a^3}{48}\frac{\partial^2 T_{1,a}}{\partial x^2} + \frac{33+11d}{48}T_{1,a}a + \frac{Aa^2(7+d)}{48D}, \tag{37}$$

and

$$T_{2,0} = \frac{a}{8D} \frac{\partial T_{1,a}}{\partial t} - \frac{a}{8} \frac{\partial^2 T_{1,a}}{\partial x^2} + \frac{15+9d}{8a} T_{1,a} + \frac{A(5+d)}{8D}. \tag{38}$$

Substitution of Eq. (38) into Eq. (10), we finally obtain two coupled advection-diffusion-reaction equations for boundary functions $T_0(x,t)$ and $T_{1,a}(x,t)$,

$$\frac{\partial T_0}{\partial t} + u_0 \frac{\partial T_0}{\partial x} = D \frac{\partial^2 T_0}{\partial x^2} + (d+1)\left[\frac{6D}{a^2}(T_w - T_0) - \frac{A}{4} - \frac{T_{1,a}}{4a} D(d+9)\right], \tag{39}$$

and

$$\frac{\partial T_{1,a}}{\partial t} = D \frac{\partial^2 T_{1,a}}{\partial x^2} + \frac{48D}{a^3}(T_w - T_0) - \frac{A}{a}(7+d) - \frac{T_{1,a} D}{a^2} 11(d+3). \tag{40}$$

Again, the original problem of solving $T(x,r,t)$ was reduced to a much simpler problem, where an analytical solution is possible. For example, the steady-state solution of $T_0$ can be found by substituting Eq. (39) into Eq. (40), where we have

$$\frac{\partial^4 T_0}{\partial x^4} - \frac{p_e}{a} \frac{\partial^3 T_0}{\partial x^3} - \frac{(39+17d)}{a^2} \frac{\partial^2 T_0}{\partial x^2} + \frac{11(3+d) p_e}{a^3} \frac{\partial T_0}{\partial x}$$

$$- \frac{18}{a^4}(d+1)(5+3d)(T_w - T_0) - \frac{3A(1+d)(5+d)}{2Da^2} = 0. \tag{41}$$

The characteristic equation (fourth order algebraic equation) for this inhomogeneous fourth order ODE is

$$r^4 - p_e r^3 - (39+17d)r^2 + 11(3+d) p_e r + 18(d+1)(5+3d) = 0. \tag{42}$$

We are only interested in the negative $r$ in order for a stable solution. Explicit solution of $r$ can only be obtained for $p_e = \infty$ and $p_e = 0$,

$$\beta_1 = -r_1 = \frac{18}{p_e} \frac{(d+1)(3d+5)}{11(d+3)} \text{ and } \beta_2 = -r_2 = \sqrt{11(d+3)} \text{ for } p_e = \infty, \tag{43}$$

$$\beta_1 = 1.5695 \text{ and } \beta_2 = 6.0446 \text{ for } p_e = 0 \text{ and } d = 0, \tag{44}$$

$$\beta_1 = 2.3935 \text{ and } \beta_2 = 7.0902 \text{ for } p_e = 0 \text{ and } d = 1, \tag{45}$$

It was noted that $\beta_1$ is strongly dependent on the Peclet number $p_e$, $\beta_2$ is less dependent on $p_e$ and approaches a constant for infinite $p_e$. The numerical solution of $\beta_1$ for $d=0$ and $d=1$ by solving the algebraic Eq. (42) is shown in Fig. 2, together with $\beta_1$ from Eq. (31) where a lower order approximation is made. The numerical solution of $\beta_2$ from Eq. (42) is shown in Fig. 3. The steady state solution of centerline temperature $T_0$ is in the form of

$$T_0 = C_1 e^{-\beta_1 \xi} + C_2 e^{-\beta_2 \xi} + C_3, \tag{46}$$

where

$$C_3 = T_w + \frac{Aa^2}{12D} \cdot \frac{5+d}{5+3d} \tag{47}$$

is the particular solution for the inhomogeneous fourth order ODE (41). The other two constants $C_1$ and $C_2$ need to be determined from inlet boundary conditions. Again, we apply the reduced boundary function method to the inlet boundary $x=0$, (Eq. (10) should be satisfied at the point $x=0$ and $r=0$).

$$\left.\frac{\partial T_0}{\partial t}\right|_{x=0} + u_0 \left.\frac{\partial T_0}{\partial x}\right|_{x=0} = D\left[\left.\frac{\partial^2 T_0}{\partial x^2}\right|_{x=0} + (d+1)T_{2,0}\big|_{x=0}\right]. \tag{48}$$

If inlet boundary has a uniform temperature $T_i$,

$$T(x=0, r, t) = T_i, \tag{49}$$

then $T_{2,0}\big|_{x=0} = 0$. From Eq. (48) we have a boundary condition for $T_0$,

$$u_0 \left.\frac{\partial T_0}{\partial x}\right|_{x=0} = D \left.\frac{\partial^2 T_0}{\partial x^2}\right|_{x=0}, \tag{50}$$

in addition to the wall boundary condition

$$T_a = T_w = T_0 + \frac{a^2}{2}T_{2,0} + \frac{a^4}{24}T_{4,0} + \frac{a^6}{120}T_{6,0}. \tag{51}$$

With the help of boundary conditions (51) and (50), we are able to determine the constants $C_1$ and $C_2$,

$$C_1 = -(C_3 - T_i)\frac{\beta_2(\beta_2 + p_e)}{\beta_2(\beta_2 + p_e) - \beta_1(\beta_1 + p_e)}, \tag{52}$$

$$C_2 = (C_3 - T_i)\frac{\beta_1(\beta_1 + p_e)}{\beta_2(\beta_2 + p_e) - \beta_1(\beta_1 + p_e)}. \tag{53}$$

With the substitution of solution of centerline temperature $T_0$ into the steady state form of Eq. (40), the solution for boundary function $T_{1,a}$ can be found,

$$T_{1,a} = D_1 e^{-\beta_1 \xi} + D_2 e^{-\beta_2 \xi} + D_3, \tag{54}$$

where

$$D_1 = \frac{48C_1}{a\left[\beta_1^2 - 11(3+d)\right]}, \tag{55}$$

$$D_2 = \frac{48C_2}{a\left[\beta_2^2 - 11(3+d)\right]}, \tag{56}$$

and

$$D_3 = -\frac{Aa}{D} \cdot \frac{(5+d)}{3(5+3d)}. \tag{57}$$

Finally, we provide a comparison of the centerline temperature solutions $T_0$ obtained using the reduced boundary function method (Eq. (30) and Eq. (46)) against the analytical solution (Eq. (58)) by use of separation of variables and neglecting the axial conduction ($p_e = \infty$) in Fig. 4. The solution is obtained for the flow confined by two flat plates (*d=0*) [19, 20],

$$\theta(\xi) = \frac{T_w - T_0}{T_w - T_i} = \sum A_n \exp(-\lambda_n^2 \xi_1), \tag{58}$$

where $\xi_1 = \xi/p_e$. The solution takes the form of infinite series with the first eight eigenvalues and coefficients shown in Table 1. Note that for small values of $\xi_1$ ($\xi_1 \to 0$) a large number of terms in series are needed for the convergence of $\theta$ and for large values of $\xi_1$ all three solutions are in good agreement with each other for $p_e = \infty$. As expected, $T_0$ obtained with high order expansion shows a better agreement with Eq. (58). However for small Peclet number $p_e = 1$, the two solutions (Eq. (30) and Eq. (46)) are quite different from the Eq. (58) showing that solution (58) is only valid for large Peclet number. However, the two solutions are in very good agreement with each other showing that high order terms do not provide significant improvement. The expansion with terms up to the fourth order (Eq. (30)) gives good results for $T_0$.

## 4. Second Type of Boundary – Heat Exchange with Surroundings

In this section, we consider heat exchange between confined flow and its surroundings. This problem is relevant to the geothermal applications, where the cold fluid is injected into hot subsurface and pumped out with thermal energy. The geometry of the problem is defined in Fig. 5. The confined flow is embedded in a medium with a temperature of $T_\infty$. It is assumed that the thickness of flow wall (*w*) is so small that heat equilibrium between external and internal wall surface can be established instantaneously, i.e the temperature at external surface equals the internal surface or the fluid temperature $T_a \equiv T(x, r = a, t)$. Newton's law of cooling is used for the heat flux between external surface and its surrounding,

$$q_o = h(T_\infty - T_a), \tag{59}$$

where $q_o$ is the heat flux at external surface and $h$ is the heat transfer coefficient between external surface and its surrounding and is assumed to be independent of the temperature difference. At the internal surface, the heat flux $q_i$ is expressed as,

$$q_i = k \frac{\partial T}{\partial r}\bigg|_{r=a} = k T_{1,a}, \tag{60}$$

Heat equilibrium between external and internal surface requires

$$q_i = q_o, \tag{61}$$

and we obtained the boundary condition for the convective heat transfer,

$$T_{1,a} = \frac{h}{k}(T_\infty - T_a). \tag{62}$$

This is a mixed Dirichlet and Neumann boundary. Since the equations derived in Section II are valid for any boundary conditions, Eq. (15) can be simplified to (with the help of the boundary condition (62))

$$-\frac{h}{k}\frac{\partial T_a}{\partial x} + \frac{\partial v}{\partial r}\bigg|_{r=a} \frac{\partial T_a}{\partial x} = D\left[-\frac{h}{k}\frac{\partial^2 T_a}{\partial x^2} + T_{3,a} + \frac{d}{a}T_{2,a} - \frac{d}{a^2}T_{1,a}\right] + \frac{2A}{a}. \tag{63}$$

We still use the expansion (Eq. (26)) with terms up to the fourth order for the temperature field $T$. The matrix form of the relationship between $T_{n,a}$ and $T_{n,0}$ ($n = 1, 2, 3$) is

$$\begin{bmatrix} T_{1,a} \\ T_{2,a} \\ T_{3,a} \end{bmatrix} = \begin{bmatrix} a & a^3/6 \\ 1 & a^2/2 \\ 0 & a \end{bmatrix} \begin{bmatrix} T_{2,0} \\ T_{4,0} \end{bmatrix}. \tag{64}$$

With the help of Eqs. (62) and (63), we can solve for the boundary functions $T_{2,a}$ and $T_{3,a}$ in terms of the boundary function $T_a$,

$$T_{2,a} = \frac{h(T_\infty - T_a)}{k} \frac{1}{a} - \frac{1}{(d+3)}\left[\frac{2A}{D} + \frac{\partial T_a}{\partial t}\frac{ah}{kD} - \frac{\partial^2 T_a}{\partial x^2}\frac{ah}{k} + \frac{\partial T_a}{\partial x}\frac{2u_0}{D}\right], \quad (65)$$

$$T_{3,a} = -\frac{1}{(d+3)}\left[\frac{6A}{Da} + \frac{\partial T_a}{\partial t}\frac{3h}{kD} - \frac{\partial^2 T_a}{\partial x^2}\frac{3h}{k} + \frac{\partial T_a}{\partial x}\frac{6u_0}{Da}\right]. \quad (66)$$

Substitution of Eq. (65) into Eq. (9) yields the governing equation for boundary function $T_a$,

$$\frac{\partial T_a}{\partial t} + \alpha \bar{u}\frac{\partial T_a}{\partial x} = D\frac{\partial^2 T_a}{\partial x^2} + \frac{D}{a^2}(1-\alpha)(1+d)(3+d)(T_\infty - T_a) + \alpha A\frac{1+d}{3+d}, \quad (67)$$

where

$$\alpha = \frac{1}{1 + \lambda/(3+d)} \quad \text{and} \quad \lambda = \frac{ah}{k}. \quad (68)$$

$\lambda$ is a dimensionless number with physical meaning similar to the Nusselt number. Obviously, the boundary function $T_a$ follows an advection-diffusion-reaction equation with an advective velocity of $v_{adv} = \alpha \bar{u}$. As expected, $v_{adv} = \bar{u}$ for a insulated boundary (zero heat flux or no heat exchange between the flowing fluid and its surroundings) where $\lambda = 0$. On the other hand, $v_{adv} = 0$ for $\lambda = \infty$, where the longitudinal conductive heat transfer is sufficiently small.

The steady-state solution of $T_a$ can be easily found by assuming that $T_i$ is the inlet temperature,

$$T_a = \left(T_i - T_\infty - \frac{Aa^2}{D}\cdot\frac{\alpha}{(1-\alpha)(3+d)^2}\right)e^{-\beta_1 \xi} + T_\infty + \frac{Aa^2}{D}\cdot\frac{\alpha}{(1-\alpha)(3+d)^2}, \quad (69)$$

where dimensionless characteristic length $\beta_1$ is dependent on both $p_e$ and $\lambda$,

$$\beta_1 = \frac{p_e \alpha}{(3+d)}\left[\sqrt{1+\frac{(1-\alpha)(d+1)(d+3)^3}{\alpha^2 p_e^2}}-1\right]. \tag{70}$$

Approximate expressions for $\beta_1$ at two limiting regimes (advection dominated and heat exchange with surroundings dominated) can be obtained as

$$\beta_1 = \frac{(1-\alpha)(d+1)(d+3)^2}{2\alpha p_e} \text{ if } \alpha \to 1 \text{ or } p_e \to \infty, \text{(advection dominated)} \tag{71}$$

$$\beta_1 = \sqrt{(1-\alpha)(d+1)(d+3)} \text{ if } \alpha \to 0 \text{ or } p_e \to 0. \text{(exchange dominated)} \tag{72}$$

Specifically, $\beta_1 = \sqrt{(d+1)(d+3)}$ for $\alpha = 0$. The plot of $\beta_1$ dependence on $p_e$ and $\alpha$ is presented in Fig. 6 for $p_e = 1, 10, 100, 1000$ increasing in the direction of the arrow. The interplay between $p_e$ and $\alpha$ determines the steady state temperature distribution.

## 5. Conclusion

The convective heat transfer was studied in great detail for a fully developed viscous flow confined by two flat plates or a circular tube. The method of solution is a reduced-boundary-function approach, where the original problem is significantly simplified by mapping the original PDE (Partial Differential Equations) onto the boundary and transforming the original problem into seeking the solutions of the unknown field at the boundary (i.e. the boundary functions). Without introducing any additional assumptions such as large Peclet number and negligible viscous dissipation, the method was applied to the convective heat transfer with two different boundary conditions. The steady state analytical solutions were shown to be in very good agreement with the series expansion method.

Table 1. The first eight eigenvalues and coefficients used in analytical solution Eq. (58).

Figure 1. A schematic representation of the convective heat transfer that is confined by two flat plates ($d=0$, left) and by a circular tube ($d=1$, right).

Figure 2. Dependence of the dimensionless parameter $\beta_1$ on the Peclet number $p_e$ for the flow confined by two flat plates ($d=0$) and by a circular tube ($d=1$) for expansions with terms up to the fourth order (Eq. (31)) and up to the sixth order (solved from Eq. (42)).

Figure 3. Dependence of dimensionless parameter $\beta_2$ (solved from Eq. (42)) on the Peclet number $p_e$ for the flow confined by two flat plates ($d=0$) and by a circular tube ($d=1$) for expansion with terms up to the sixth order.

Figure 4. Comparison of dependence of dimensionless centerline temperature $\theta$ on the axial position $\xi_1$ for Eqs. (30) (46) and (58).

Figure 5. A schematic representation of the convective heat transfer with heat exchange with surroundings.

Figure 6. Dependence of dimensionless parameter $\beta_1$ on the Peclet number $p_e$ and parameter $\alpha$ for the flow confined by two flat plates ($d=0$) and by a circular tube ($d=1$).

Table. 1

| n | $\lambda_n$ | $A_n$ |
|---|---------|---------|
| 0 | 1.6816  | 1.2005  |
| 1 | 5.6699  | -0.2991 |
| 2 | 9.6683  | 0.1608  |
| 3 | 13.6677 | -0.1074 |
| 4 | 17.6674 | 0.0796  |
| 5 | 21.6672 | -0.0628 |
| 6 | 25.6671 | 0.0512  |
| 7 | 29.6670 | -0.0483 |

Fig. 1

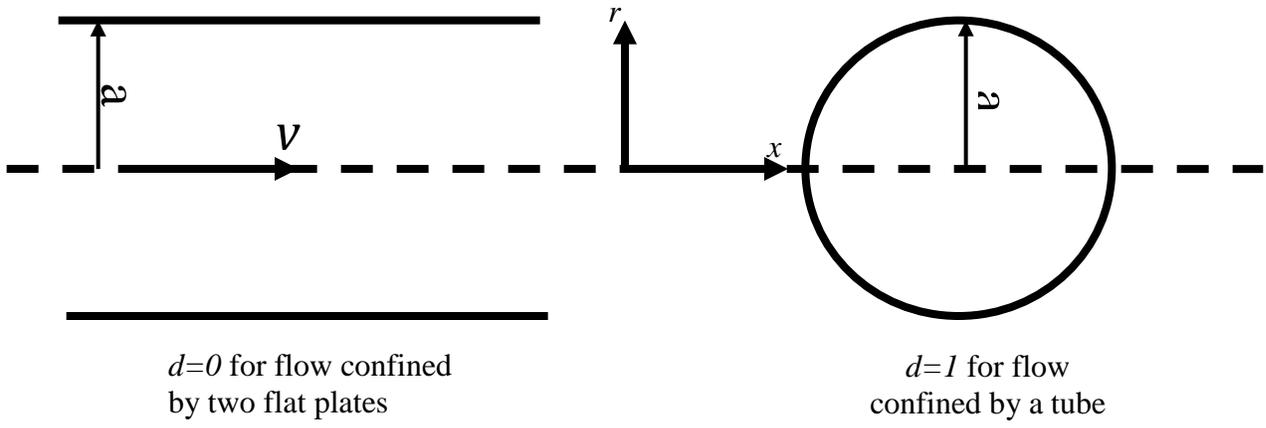

*d=0* for flow confined by two flat plates

*d=1* for flow confined by a tube

Fig. 2

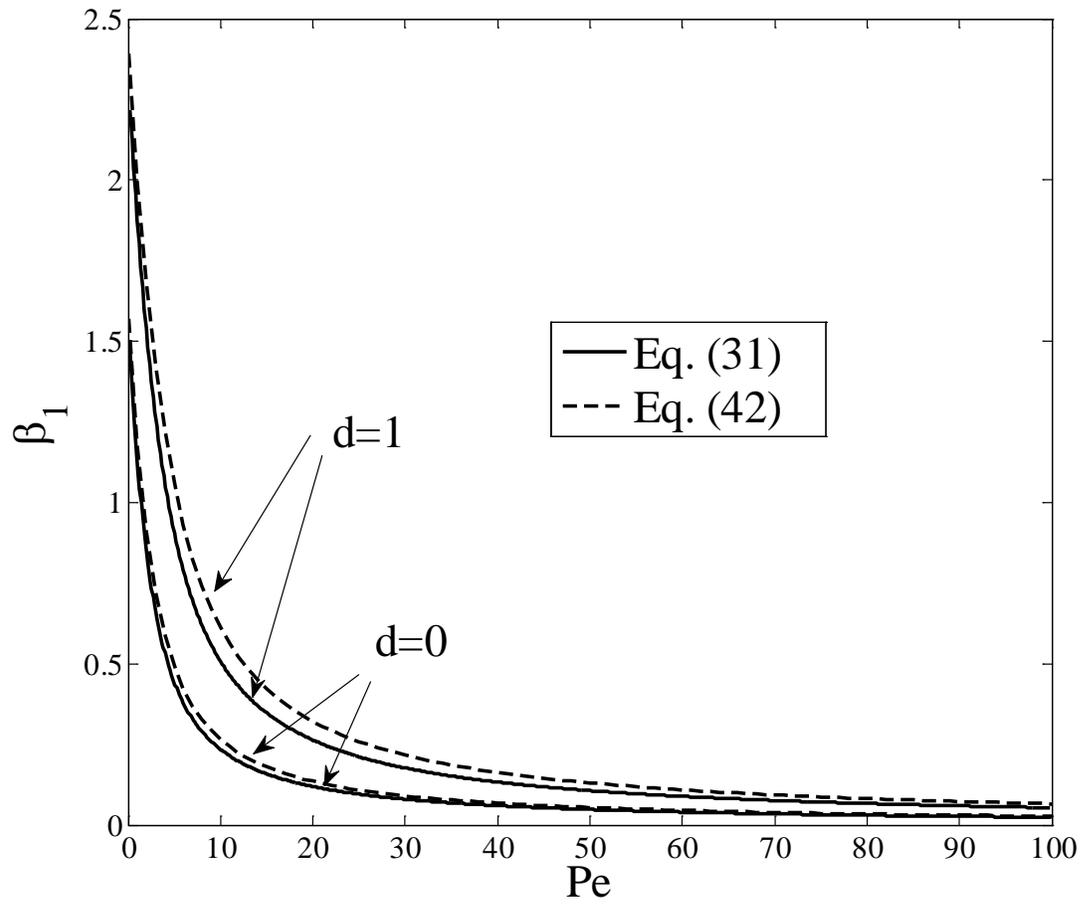

Fig. 3

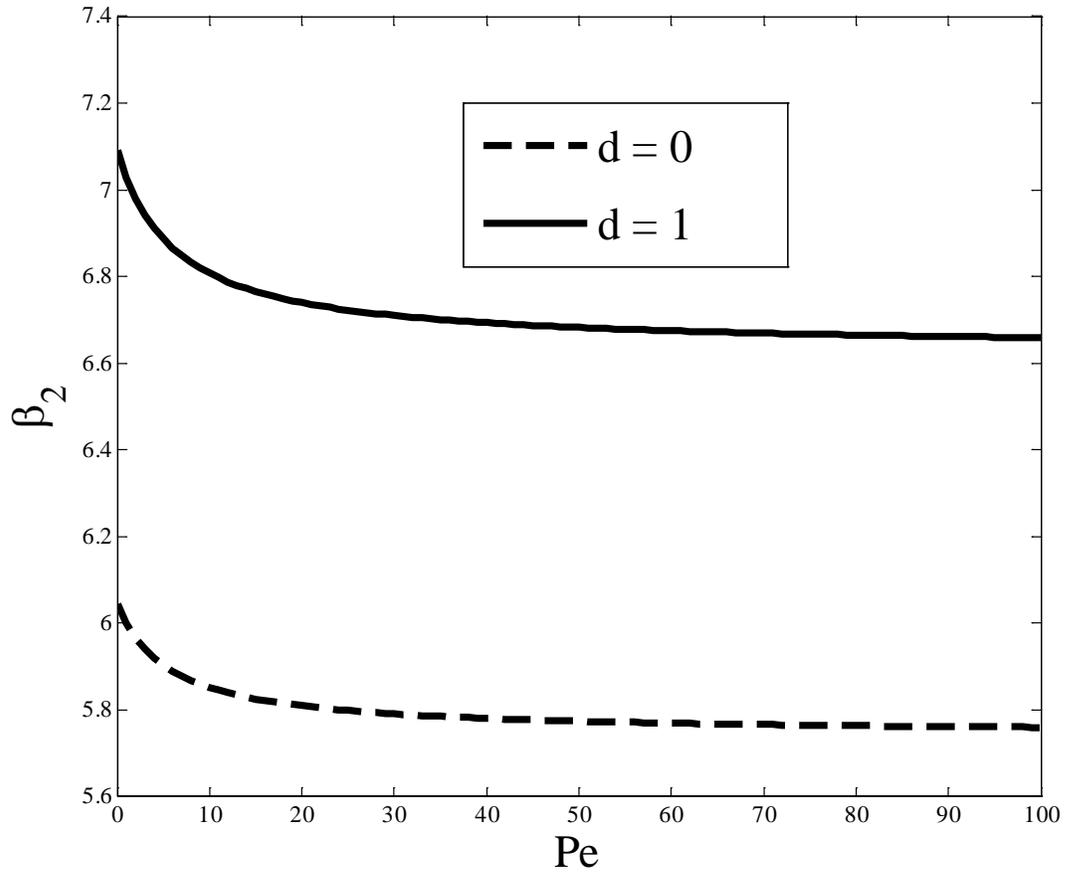

Fig.4

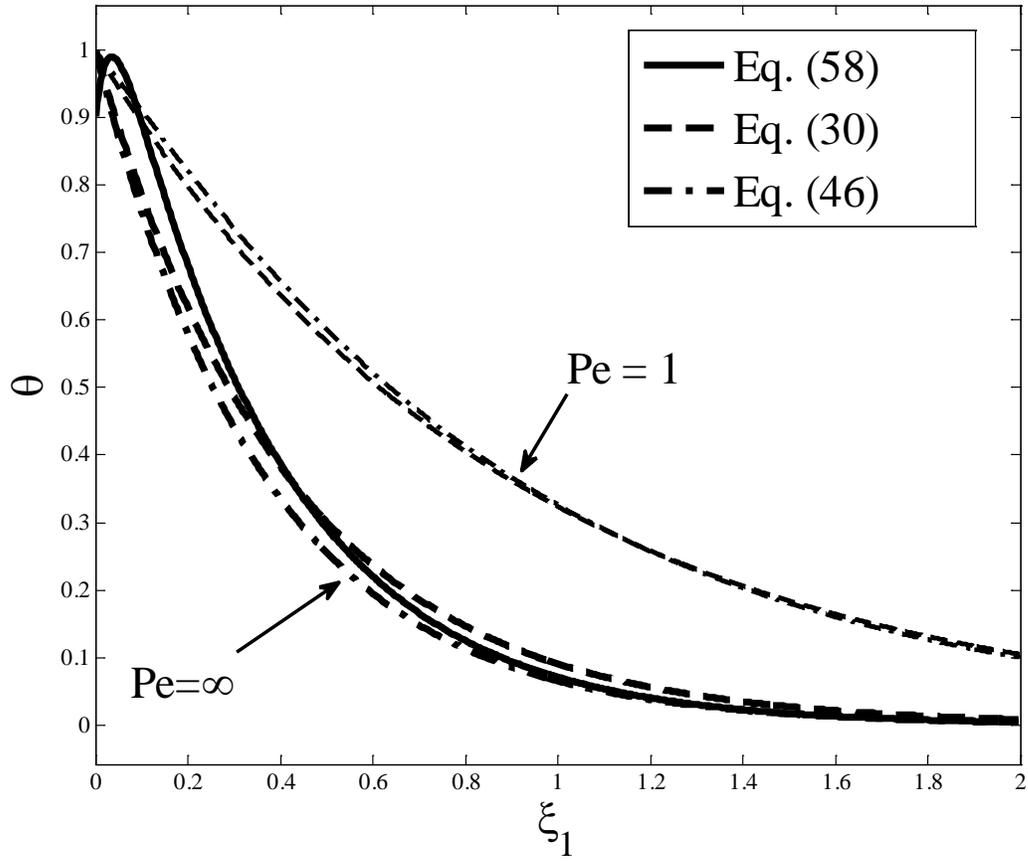

Fig. 5

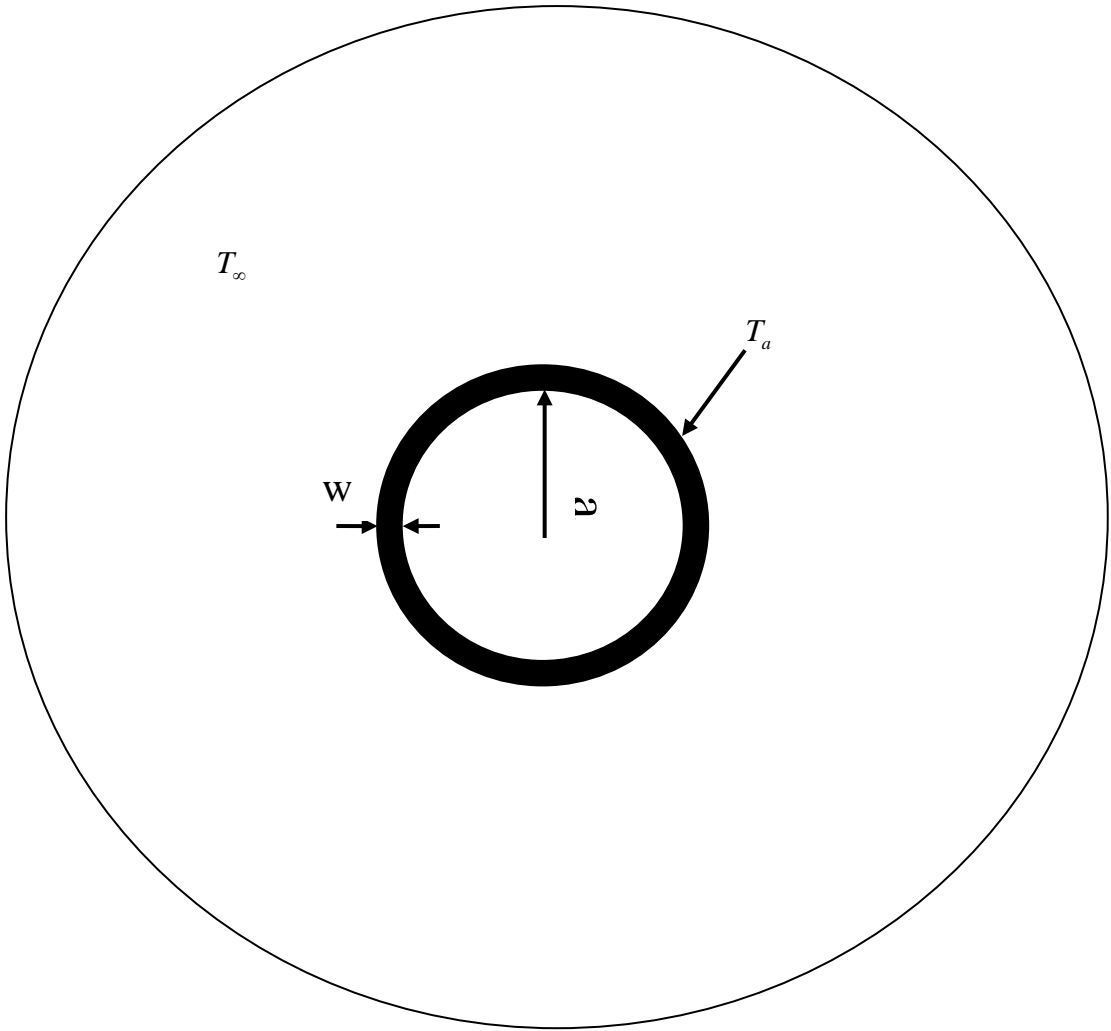

Fig. 6.

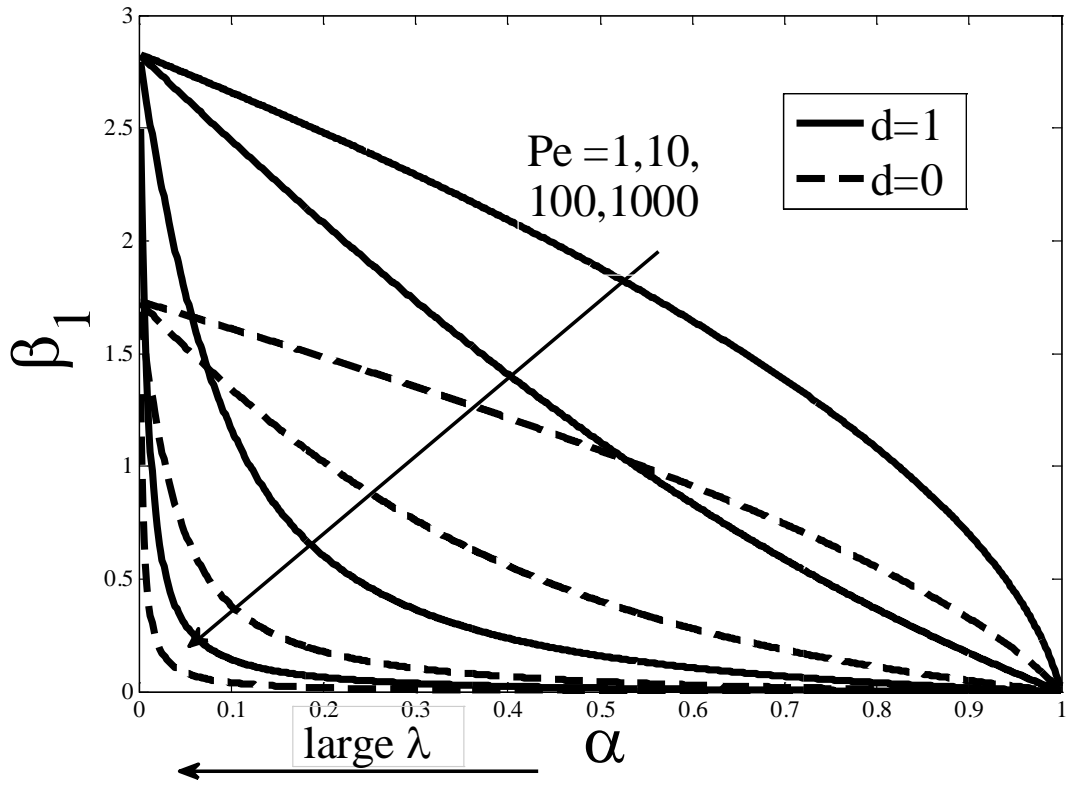